\documentclass[11pt]{article}
\usepackage{epstopdf}
\usepackage{color}
\usepackage{subfigure}
\usepackage{amsmath}
\usepackage{amssymb}
\usepackage{graphicx,color}
\usepackage{cite}
\usepackage{enumerate}
\usepackage{amsthm}% theorems and proofs
\usepackage{amsfonts,mathrsfs}
\usepackage{geometry}
\usepackage{ifthen}

\begin{document}

\title{Coherence and complementarity based on modified generalized skew information}

\vskip0.1in
\author{\small Zhaoqi Wu$^{1,4}$, Lin Zhang$^{2,4}$\thanks{Corresponding author. E-mail:
godyalin@163.com;linzhang@mis.mpg.de}, Shao-Ming
Fei\thanks{Corresponding author. E-mail: feishm@cnu.edu.cn}
$^{3,4}$,
Xianqing Li-Jost$^{4}$\\
{\small\it  1. Department of Mathematics, Nanchang University, Nanchang 330031, P R China} \\
{\small\it  2. Institute of Mathematics, Hangzhou Dianzi University, Hangzhou 310018, P R China}\\
{\small\it  3. School of Mathematical Sciences, Capital Normal University, Beijing 100048, P R China}\\
{\small\it  4. Max-Planck-Institute for Mathematics in the Sciences, 04103 Leipzig, Germany}}

\date{}
\maketitle

\noindent {\bf Abstract} {\small } We introduce modified generalized
Wigner-Yanase-Dyson (MGWYD) skew information and modified weighted
generalized Wigner-Yanase-Dyson (MWGWYD) skew information. By
revisiting state-channel interaction based on MGWYD skew
information, a family of coherence measures with respect to quantum
channels is proposed. Furthermore, explicit analytical expressions
of these coherence measures of qubit states are derived with respect
to different quantum channels. Moreover, complementarity relations
based on MGWYD skew information and MWGWYD skew information are also
presented. Specifically, the conservation relations are
investigated, while two interpretations of them including
symmetry-asymmetry complementarity and wave-particle duality have
been proposed.

\vskip 0.1 in

\noindent PACS numbers: 03.65.Ud, 03.67.-a, 03.75.Gg

\noindent {\bf Key Words}: {\small } Coherence; complementarity; modified generalized Wigner-Yanase-Dyson skew information; modified weighted generalized Wigner-Yanase-Dyson skew information
\vskip0.2in

\noindent {\bf 1. Introduction}

Originating from the superposition principle, quantum coherence is a
characteristic feature of quantum mechanics. Despite its wide
applications in superconductivity, quantum thermodynamics and
biological processes, the quantification of quantum coherence from a
resource-theoretic perspective was initiated only recently
\cite{TB}. Since then the study on quantum coherence has attracted
much attention in recent years. Various kinds of coherence measures
such as relative entropy of coherence, $l_1$ norm of coherence,
intrinsic randomness of measurement, robustness of coherence,
averaged coherence, max-relative entropy of coherence, modified
trace distance, skew information, Hellinger distance, affinity
distance and generalized $\alpha$-$z$-relative R\'enyi entropy have
been proposed to quantify quantum coherence
\cite{XY,AW,CN,LZ1,LZ2,BC,KB1,CY,LUO8,LUO9,LUO10,ZXJ,CX1,XNZ}.
Coherence-generating power of quantum channels has also been
investigated \cite{PZ1,PZ2,LZ3}.

On the other hand, the relationships between quantum coherence and
other quantum resources such as quantum entanglement, quantum
discord and asymmetry \cite{AS1,CR,JM,EC,HJZ,MP,IM,KB2} have been
extensively studied. It has been shown that quantum coherence and
asymmetry cannot be broadcasted \cite{Muller,Marvian}. Quantum
coherence is also tightly related to the optimization of quantum
observables \cite{KCT}. Recently, a regime of defining coherence
measures by making use of POVMs has been put forward \cite{BKB}.
Utilizing the concept of resource destroying maps, the authors in
\cite{TEZP} have established a framework of coherence theory on the
level of quantum operations.

Complementarity is another important quantum feature which has been
extensively studied since the advent of quantum mechanics. It is
well known that the Bohr's complementarity principle plays an
indispensable role in the basic theory of quantum mechanics in the
early days \cite{NB}. This was manifested in wave-particle duality
and uncertainty relations by many authors
\cite{WH1,GBE,SD,XP,JST,FS,PJC1,PJC2,PJC3,EB}.

Recently, by decomposing the state-channel interaction into a
symmetric and an asymmetric part, the authors in \cite{LUO11}
formalized a quantitative symmetry-asymmetry complementarity
relation. The asymmetric part is given based on a modified version
of Wigner-Yanase skew information, in which a Hermitian operator is
replaced by a bounded linear operator (not-necessarily-Hermitian),
which can be interpreted as a measure of coherence with respect to a
quantum channel. However, as a desired property of a quantum
coherence measure, the strong monotonicity of this quantity is not
proved in \cite{LUO11}. In \cite{LWJ}, Li provided an alternative
proof of the monotonicity based on the modified skew information via
operator algebra approach, and derived the strong monotonicity.

The generalized Wigner-Yanase-Dyson skew information
with parameters $\alpha$ and $\beta$ has been introduced in \cite{CL}.
It would be interesting if such generalized Wigner-Yanase-Dyson skew information
could be utilized to characterize quantum coherence, general state-channel interactions
and complementarity relations among the coherence measures.
In this work, we define correspondingly the modified generalized Wigner-Yanase-Dyson
(MGWYD) skew information and the modified weighted generalized
Wigner-Yanase-Dyson (MWGWYD) skew information. Based on the MGWYD
skew information and MWGWYD skew information, we study the
state-channel interactions. It is shown that the asymmetric part of
the generalized state-channel couplings can be regarded as a family
of coherence of states with respect to a channel. Complementarity
relations based on MGWYD skew information and MWGWYD skew
information are also derived with physical interpretations. Some
concluding remarks are given in Section 6. Our results shed new
light on the study of coherence, and give rise to a basic framework for
quantitatively addressing symmetry-asymmetry complementarity.

\vskip0.1in

\noindent {\bf 2. Preliminaries}

\vskip0.1in

Let $\mathcal{H}$ be a $d$-dimensional Hilbert space, and
$\mathcal{B(H)}$, $\mathcal{S(H)}$ and $\mathcal{D(H)}$ the set of
all bounded linear operators, Hermitian operators and density
operators on $\mathcal{H}$, respectively. Usually, a state and a
channel are mathematically described by a density operator (positive
operator of trace $1$) and a completely positive trace preserving
(CPTP) map, respectively \cite{NC}. Nevertheless, in this
paper when we discuss state-channel interactions, a channel is assumed to be a completely positive trace
nonincreasing map, while a quantum operation is assumed to be a
completely positive trace preserving (CPTP) map, in accord with
Ref. \cite{LUO11}.

Fix an orthonormal basis $\{|i\rangle\}^d_{i=1}$ of a
$d$-dimensional Hilbert space $\mathcal{H}$. The density operators
which are diagonal in this basis are called incoherent states and
the set of all incoherent states is denoted by $\mathcal{I}$, i.e.,
$$\mathcal{I}=\{\delta\in \mathcal{D(H)}|\delta=\sum_{i}p_i|i\rangle\langle i|,~p_i\geq 0,~\sum_{i}p_i=1\}.$$
Let $\Lambda$ be a completely positive trace preserving (CPTP) map
$$\Lambda(\rho)=\sum_{n}K_n\rho K_n^\dag,$$
where $K_n$ are Kraus operators satisfying $\sum_{n}K_n^\dag
K_n=I_{d}$ with $I_d$ the identity operator. $K_n$ are called
incoherent Kraus operators if $K_n^\dag \mathcal{I}K_n\in
\mathcal{I}$ for all $n$, and the corresponding
$\Lambda$ is called an incoherent operation.

In \cite{TB}, the authors proposed the conditions that a well-defined coherence measure $C$
should satisfy:

$(C_1)$ (Faithfulness) $C(\rho)\geq 0$ and $C(\rho)=0$ iff $\rho$ is
incoherent.

$(C_2)$ (Monotonicity) $C(\Lambda(\rho))\leq C(\rho)$ for any
incoherent operation $\Lambda$.

$(C_3)$ (Convexity) $C(\cdot)$ is a convex function of $\rho$, i.e.,
$$\sum_{n}p_nC(\rho_n)\geq C(\sum_{n}p_n\rho_n),$$
where $p_n\geq 0,~ \sum_{n}p_n=1$.

$(C_4)$ (Strong monotonicity) $C(\cdot)$ does not increase on
average under selective incoherent operations, i.e.,
$$C(\rho)\geq \sum_{n}p_nC(\varrho_n),$$
where $p_n=\mathrm{Tr}(K_n\rho K_n^\dag)$ are probabilities and
$\varrho_n=\frac{K_n\rho K_n^\dag}{p_n}$ are the post-measurement states,
$K_n$ are incoherent Kraus operators.

Now, we recall the concepts of different kinds of skew information.
For a density operator $\rho\in \mathcal{D(H)}$ and an observable
$A\in \mathcal{S(H)}$, the {\it Wigner-Yanase} (WY) skew information
\cite{WY} is defined by
\begin{equation}\label{eq1}
I_{\rho}(A)=-\frac{1}{2}\mathrm{Tr}([\rho^{\frac{1}{2}},A]^2),
\end{equation}
where $[X,Y]:=XY-YX$ is the commutator of $X$ and $Y$. A more
general quantity was suggested by Dyson,
\begin{equation}\label{eq2}
I_{\rho}^{\alpha}(A)=-\frac{1}{2}\mathrm{Tr}([\rho^{\alpha},A][\rho^{1-\alpha},A]),\,\,~0\leq
\alpha \leq 1,
\end{equation}
which is now called the {\it Wigner-Yanase-Dyson} (WYD) skew
information. The quantity in Eq. (\ref{eq2}) was further generalized
to \cite{CL}
\begin{equation}\label{eq3}
I_{\rho}^{\alpha,\beta}(A)=-\frac{1}{2}\mathrm{Tr}([\rho^\alpha,
A][\rho^\beta, A]\rho^{1-\alpha-\beta}),~~~\alpha,\beta\geq
0,~\alpha+\beta\leq 1,
\end{equation}
which is termed as {\it generalized Wigner-Yanase-Dyson} (GWYD)
skew information. It is easy to see that when $\alpha+\beta=1$, Eq.
(\ref{eq3}) reduces to Eq. (\ref{eq2}), and Eq. (\ref{eq2}) reduces
to Eq. (\ref{eq1}) when $\alpha=\frac{1}{2}$.

Another generalization of WYD skew information was given in \cite{FURU1}
\begin{equation}\label{eq4}
K_{\rho}^{\alpha}(A)=-\frac{1}{2}\mathrm{Tr}\left(\left[\frac{\rho^\alpha+\rho^{1-\alpha}}{2},
A_0\right]^2\right) ~,\,\,0\leq \alpha \leq 1,
\end{equation}
where $A_0=A-\mathrm{Tr}(\rho A)I$. We call $K_{\rho}^{\alpha}(A)$
the {\it weighted Wigner-Yanase-Dyson skew information} in the
following. Noting that $I_{\rho}(A)=I_{\rho}(A_0)$ when $\alpha=\frac{1}{2}$, one sees
that Eq. (\ref{eq4}) also reduces to Eq. \eqref{eq1} in this case.

Remarkable properties of these quantities and their applications in
quantum information theory have been revealed and explored during
the past few years
\cite{LUO1,LUO3,LUO4,LUO5,LUO6,LUO7,FURU2,FURU3,BC2,MCF}.
Nevertheless, quantum gates \cite{NC}, generalized quantum gates
\cite{LGL}, the Kraus operators of a quantum channel \cite{NC} and
many other operators are not necessarily Hermitian. Therefore, it is
natural to consider the corresponding definitions of the different
types of the skew information mentioned above for pseudo-Hermitian
and/or {\cal PT}-symmetric quantum mechanics
\cite{Bender98,Makris-PRL,Guo-PRL,Ruter-NP,Chang-NP,Tang-NP}.

For a density operator $\rho\in \mathcal{D(H)}$ and an operator
$A\in \mathcal{B(H)}$ (not necessarily Hermitian), a generalization
of the quantity in Eq. (\ref{eq1}) is defined by \cite{DOU1}
\begin{equation}\label{eq5}
|I_{\rho}|(A)=-\frac{1}{2}\mathrm{Tr}([\rho^{\frac{1}{2}},A^\dag][\rho^{\frac{1}{2}},A]),
\end{equation}
which we refer to {\it modified Wigner-Yanase} (MWY) skew
information.
Similarly, a generalization of the quantity in Eq. (\ref{eq2}) is
defined by \cite{DOU2}
\begin{equation}\label{eq6}
|I_{\rho}^{\alpha}|(A)=-\frac{1}{2}\mathrm{Tr}([\rho^{\alpha},A^\dag][\rho^{1-\alpha},A]),\,\,~0\leq
\alpha \leq 1,
\end{equation}
for any $A\in \mathcal{B(H)}$ and $\rho\in \mathcal{D(H)}$, which we
call {\it modified Wigner-Yanase-Dyson} (MWYD) skew information.
A generalization of the quantity in Eq. (\ref{eq4}) is given by
\cite{CZL2}
\begin{equation}\label{eq7}
|K_{\rho}^{\alpha}|(A)=-\frac{1}{2}\mathrm{Tr}\left(\left[\frac{\rho^\alpha+\rho^{1-\alpha}}{2},
A_0^\dag\right]\left[\frac{\rho^\alpha+\rho^{1-\alpha}}{2},
A_0\right]\right) ,\,\,~0\leq \alpha \leq 1,
\end{equation}
for any $A\in L^2(H)$ and $\rho\in D(H)$, which we call {\it
modified weighted Wigner-Yanase-Dyson} (MWWYD) skew information. The
related quantity $|L_{\rho}^{\alpha}|(A)$ is defined by replacing the
commutators in Eq. (\ref{eq7}) by anti-commutators.

In addition, a {\it Schatten $p$-norm} \cite{RB} is defined as
$$\|A\|_{p}=\left[\sum_{j=1}^n(s_j(A))^p\right]^{1/p},$$
where $s_j(A)$ denotes the singular value of $A$. When $p=2$, it is
called a {\it Hilbert-Schmidt norm} \cite{RB}.
Note that the class of Schatten $p$-norms is a special type of
unitarily invariant norms \cite{RB} satisfying
$|||UAV|||=|||A|||\,\,\,\makebox{for all}\,\, A\in M(n)
\,\,\makebox{and}\,\,U,V\in U(n)$, where $M(n)$ denotes the set of
$n\times n$ matrices and $U(n)$ the unitary group in $M(n)$.

Recently, Luo et al. defined the following quantity for any operator
$K\in \mathcal{B(H)}$ and state $\rho\in \mathcal{D(H)}$
\cite{LUO11},
\begin{equation}\label{eq8}
I({\rho},K)=\mathrm{Tr}([\rho^{\frac{1}{2}},K]^\dag[\rho^{\frac{1}{2}},K])=\|[\rho^{\frac{1}{2}},K]\|_2^2,
\end{equation}
where $\|X\|_2^2=\mathrm{Tr}(X^\dag X)$ is the Hilbert-Schmidt norm,
$[X,Y]=\frac{1}{2}(XY-YX)$ is the commutator. This quantity is in
fact the one defined in \cite{DOU1} (up to a constant factor). For
the sake of convenience, we call it {\it modified Wigner-Yanase
(MWY) skew information} in the following.
Similarly, the following quantity was defined as a measure to
quantify the symmetry between $\rho$ and $K$,
\begin{equation}\label{eq9}
J({\rho},K)=\mathrm{Tr}(\{\rho^{\frac{1}{2}},K\}
\{\rho^{\frac{1}{2}},K\}^\dag)=\|\{\rho^{\frac{1}{2}},K\}\|_2^2,
\end{equation}
where $\{X,Y\}=\frac{1}{2}(XY+YX)$ is the anti-commutator.

Any quantum channel (completely positive trace nonincreasing map)
has the following Kraus representation
\begin{equation}\label{eq10}
\Phi(\rho)=\sum_{i}K_i\rho K_i^\dag,
\end{equation}
while the dual channel can be written as
\begin{equation}\label{eq11}
\Phi^{\dag}(X)=\sum_{i}K_i^\dag X K_i,
\end{equation}
where $X$ is any non-negative operator.

With respect to Eqs. (\ref{eq8}) and (\ref{eq9}), for a quantum
channel in the form of Eq. (\ref{eq10}) , it is defined that
\begin{equation}\label{eq12}
I({\rho},\Phi)=\sum_{i}I(\rho,K_i)
\end{equation}
and
\begin{equation}\label{eq13}
J({\rho},\Phi)=\sum_{i}J(\rho,K_i).
\end{equation}
For any operator $K\in \mathcal{B(H)}$ and state $\rho\in
\mathcal{D(H)}$, the following quantity \cite{LWJ} has been derived
from the unified entropy
\begin{equation}\label{eq14}
I^{\alpha}(\rho,K)=\mathrm{Tr}([\rho^{\alpha},K]^\dag[\rho^{1-\alpha},K]),~~0\leq
\alpha \leq 1.
\end{equation}
Note that this quantity is in fact the one defined in \cite{DOU2}
(up to a constant factor). We shall call it {\it modified
Wigner-Yanase-Dyson (MWYD) skew information} in the following.
$J^{\alpha}(\rho,K)$ was defined by using anti-commutator in Eq.
(\ref{eq12}). Also, $I^{\alpha}(\rho,\Phi)$ and
$J^{\alpha}(\rho,\Phi)$ have been proposed in a similar manner with
the Kraus representation of a quantum channel.

\vskip0.1in

\noindent {\bf 3. State-channel interactions based on modified
generalized skew information}

\vskip0.1in

In this section, we will study state-channel interaction based on
modified generalized skew information. We first define the {\it
modified generalized Wigner-Yanase-Dyson (MGWYD) skew information}
for any operator $K\in \mathcal{B(H)}$ and state $\rho\in
\mathcal{D(H)}$,
\begin{equation}\label{eq15}
I^{\alpha,\beta}(\rho,K)=\mathrm{Tr}([\rho^\alpha, K]^\dag[\rho^\beta, K]\rho^{1-\alpha-\beta}),~~\alpha,\beta\geq 0,~\alpha+\beta\leq 1,
\end{equation}
and the related quantity $J^{\alpha,\beta}(\rho,K)$,
\begin{equation}\label{eq16}
J^{\alpha,\beta}(\rho,K)=\mathrm{Tr}(\{\rho^\alpha, K\}^\dag\{\rho^\beta, K\}\rho^{1-\alpha-\beta}),~~\alpha,\beta\geq 0,~\alpha+\beta\leq 1.
\end{equation}

Note that when $\alpha+\beta=1$, $I^{\alpha,\beta}(\rho,K)$ and $J^{\alpha,\beta}(\rho,K)$ reduce to $I^{\alpha}(\rho,K)$ and $J^{\alpha}(\rho,K)$, respectively. Furthermore,
corresponding to the map $\Phi(\rho)=\sum_{i}K_i\rho K_i^\dag$, we define
\begin{equation}\label{eq17}
I^{\alpha,\beta}({\rho},\Phi)=\sum_{i}I^{\alpha,\beta}(\rho,K_i),~~\alpha,\beta\geq 0,~\alpha+\beta\leq 1
\end{equation}
and
\begin{equation}\label{eq18}
J^{\alpha,\beta}({\rho},\Phi)=\sum_{i}J^{\alpha,\beta}(\rho,K_i),~~\alpha,\beta\geq 0,~\alpha+\beta\leq 1.
\end{equation}

On the other hand, we define the {\it modified weighted generalized Wigner-Yanase-Dyson (MWGWYD) skew information} for any operator $K\in \mathcal{B(H)}$ and state $\rho\in \mathcal{D(H)}$,
\begin{equation}\label{eq19}
V^{\alpha,\beta}(\rho,
K)=\mathrm{Tr}\left(\left[\frac{\rho^\alpha+\rho^{\beta}}{2},
K\right]^\dag\left[\frac{\rho^\alpha+\rho^{\beta}}{2},
K\right]\rho^{1-\alpha-\beta}\right),~~\alpha,\beta\geq
0,~\alpha+\beta\leq 1,
\end{equation}
with the related quantity
\begin{equation}\label{eq20}
W^{\alpha,\beta}(\rho,
K)=\mathrm{Tr}\left(\left\{\frac{\rho^\alpha+\rho^{\beta}}{2},K\right\}^\dag\left\{\frac{\rho^\alpha+\rho^{\beta}}{2},
K\right\}\rho^{1-\alpha-\beta}\right),~~\alpha,\beta\geq
0,~\alpha+\beta\leq 1.
\end{equation}
Obviously, when $\alpha+\beta=1$, $V^{\alpha,\beta}(\rho,K)$ and
$W^{\alpha,\beta}(\rho,K)$ reduce to $|K_{\rho}^{\alpha}|(A)$ and
$|L_{\rho}^{\alpha}|(A)$ defined in \cite{CZL2} (up to a constant
factor), respectively. With respect to the map
$\Phi(\rho)=\sum_{i}K_i\rho K_i^\dag$, we define
\begin{equation}\label{eq21}
V^{\alpha,\beta}({\rho},\Phi)=\sum_{i}V^{\alpha,\beta}(\rho,K_i),~~\alpha,\beta\geq 0,~\alpha+\beta\leq 1
\end{equation}
and
\begin{equation}\label{eq22}
W^{\alpha,\beta}({\rho},\Phi)=\sum_{i}W^{\alpha,\beta}(\rho,K_i),~~\alpha,\beta\geq
0,~\alpha+\beta\leq 1.
\end{equation}

{\bf Remark 1} As noted in \cite{LUO11} and \cite{LWJ}, we can also
prove that the quantities $I^{\alpha,\beta}({\rho},\Phi)$,
$J^{\alpha,\beta}({\rho},\Phi)$, $V^{\alpha,\beta}({\rho},\Phi)$ and
$W^{\alpha,\beta}({\rho},\Phi)$ are independent of the choice of the
Kraus operators of $\Phi$, which guarantees that the quantities
given by Eqs. (\ref{eq17}), (\ref{eq18}), (\ref{eq21}) and
(\ref{eq22}) are all well-defined.

From the above definitions, we have the following results
of $I^{\alpha,\beta}(\rho,K)$ (the $W^{\alpha,\beta}(\rho,K)$ admits similar properties).

{\bf Proposition 1} For $\alpha,\beta\geq 0$, $\alpha+\beta\leq 1$, $K\in \mathcal{B(H)}$ and $\rho\in \mathcal{D(H)}$, it holds that

(i) $I^{\alpha,\beta}(\rho,K)=I^{\beta,\alpha}(\rho,K)$;

(ii) If $\alpha,\beta\in [0,1]$, $\alpha+2\beta\leq 1$ and $2\alpha+\beta\leq 1$, then $I^{\alpha,\beta}(\rho,K)$ is convex in $\rho$. In particular, $I^{\alpha}(\rho,K)$ is convex in $\rho$ for $0\leq \alpha \leq 1$.

{\bf Proof.} (i) follows immediately from the definition. It follows from \cite{LUO9} that (ii) holds when $K\in \mathcal{S(H)}$. It can be seen that (ii) also holds for any $K\in \mathcal{B(H)}$ from the proof in \cite{LUO6}.

\vskip0.1in

\noindent {\bf 4. A family of coherence measures of a state with
respect to a channel}

\vskip0.1in

In this section, we demonstrate that $I^{\alpha,\beta}({\rho},\Phi)$
could be viewed as a bona fide family of coherence measures of
$\rho$ with respect to a channel $\Phi$ under certain conditions.
We first prove the following properties of $I^{\alpha,\beta}({\rho},\Phi)$.

{\bf Theorem 1} For channel $\Phi(\rho)=\sum_{j}K_j\rho K_j^\dag$,
the quantity $I^{\alpha,\beta}({\rho},\Phi)$ defined in Eq.
(\ref{eq17}) for $\alpha,\beta\geq 0$ with $\alpha+\beta\leq 1$ has
the following properties:

(i) $I^{\alpha,\beta}(\rho,\Phi)\geq 0$, with the equality holding if and only if $\Phi^\dag(\rho^{\alpha})=\rho^{\alpha}$, $\Phi^\dag(\rho^{\beta})=\rho^{\beta}$ and $\Phi^\dag(\rho^{\alpha+\beta})=\rho^{\alpha+\beta}$.

(ii) If $\alpha,\beta\in [0,1]$, $\alpha+2\beta\leq 1$ and
$2\alpha+\beta\leq 1$, then $I^{\alpha,\beta}(\rho,\Phi)$ is convex
in $\rho$.

(iii) (Ancillary independence) $I^{\alpha,\beta}(\rho^a\otimes \rho^b,\Phi^a\otimes \mathbf{1}^b)=I^{\alpha,\beta}(\rho^a,\Phi^a)$, where $\rho^a$ and $\rho^b$ are any states of systems $a$ and $b$, respectively, and $\mathbf{1}^b$ is the identity channel on system $b$.

(iv) (Monotonicity) If a channel $\mathcal{E}$ admits the representation $\mathcal{E}(\rho)=\sum_{i}E_i\rho E_i^\dag$, then
$$
I^{\alpha,\beta}(\mathcal{E}(\rho),\Phi)\geq
I^{\alpha,\beta}(\rho,\Phi),
$$
provided that one of the following two conditions is satisfied for
all $i$:

(1) $\mathcal{E}^\dag(K_i)=K_i$, $\mathcal{E}^\dag(K_i^\dag K_i)=K_i^\dag K_i$, $\mathcal{E}^\dag(K_i K_i^\dag)=K_i K_i^\dag$ and $[\rho^{1-\alpha-\beta},K_i]=0$;

(2) $\mathcal{E}^\dag(K_i)=K_i$, $\mathcal{E}^\dag(K_i^\dag
K_i)=K_i^\dag K_i$ and $[\rho^{\alpha+\beta},K_i]=0$.

(v) (Strong monotonicity) If a channel $\mathcal{E}$ admits the representation $\mathcal{E}(\rho)=\sum_{i}E_i\rho E_i^\dag$, then
$$
\sum_{i}p_iI^{\alpha,\beta}(\rho_i,\Phi)\geq I^{\alpha,\beta}(\rho,\Phi),
$$
where $p_i=\mathrm{Tr}E_i\rho E_i^\dag$ and $\rho_i=E_i\rho
E_i^\dag/p_i$, provided that condition (1) or (2) in item (iv) is
satisfied for all $i$ and $\alpha,\beta>0$ with $\alpha+\beta\leq
1$.

{\bf Proof.} (i) and (ii) can be verified easily from the
definitions and the property (ii) in Proposition 1. Direct
calculation shows that
\begin{eqnarray*}
&&4I^{\alpha,\beta}(\rho^a\otimes \rho^b,\Phi^a\otimes \mathbf{1}^b)
\\&=&1-\mathrm{Tr}[(\rho^a\otimes \rho^b)^{1-\alpha}(\Phi^{a\dag}\otimes \mathbf{1}^b)((\rho^a\otimes \rho^b)^{\alpha})]
-\mathrm{Tr}[(\rho^a\otimes \rho^b)^{1-\beta}(\Phi^{a\dag}\otimes
\mathbf{1}^b)((\rho^a\otimes \rho^b)^{\beta})]\notag
\\&& -\mathrm{Tr}[(\rho^a\otimes \rho^b)^{1-\alpha-\beta}(\Phi^{a\dag}\otimes \mathbf{1}^b)((\rho^a\otimes \rho^b)^{\alpha+\beta})]
\\&=&1-\mathrm{Tr}[(\rho^a)^{1-\alpha}\otimes (\rho^b)^{1-\alpha}(\Phi^{a\dag}((\rho^a)^{\alpha})\otimes (\rho^b)^{\alpha})]
-\mathrm{Tr}[(\rho^a)^{1-\beta}\otimes
(\rho^b)^{1-\beta}(\Phi^{a\dag}((\rho^a)^{\beta})\otimes
(\rho^b)^{\beta})]\notag
\\&&+\mathrm{Tr}[(\rho^a)^{1-\alpha-\beta}\otimes (\rho^b)^{1-\alpha-\beta}(\Phi^{a\dag}((\rho^a)^{\alpha+\beta})\otimes (\rho^b)^{\alpha+\beta})]
\\&=& 1-\mathrm{Tr}[(\rho^a)^{1-\alpha}(\Phi^{a\dag}((\rho^a)^{\alpha}))]-\mathrm{Tr}[(\rho^a)^{1-\beta}(\Phi^{a\dag}((\rho^a)^{\beta}))]
+\mathrm{Tr}[(\rho^a)^{1-\alpha-\beta}(\Phi^{a\dag}((\rho^a)^{\alpha+\beta}))]
\\&=&4I^{\alpha,\beta}(\rho^a,\Phi^a).
\end{eqnarray*}
Hence item (iii) holds.

It follows from Eq. (\ref{eq15}) that
$$
I^{\alpha,\beta}(\rho,K_i)=\frac{1}{4}[\mathrm{Tr}(\rho^{1-\alpha-\beta}K_i^\dag \rho^{\alpha+\beta}K_i)+\mathrm{Tr}(\rho K_i^\dag K_i)-\mathrm{Tr}(\rho^{1-\alpha}K_i^\dag \rho^{\alpha}K_i)-\mathrm{Tr}(\rho^{1-\beta}K_i^\dag \rho^{\beta}K_i)].
$$
If $[\rho^{1-\alpha-\beta},K_i]=0$, the above equation becomes
$$
I^{\alpha,\beta}(\rho,K_i)=\frac{1}{4}[\mathrm{Tr}(\rho K_i K_i^\dag)+\mathrm{Tr}(\rho K_i^\dag K_i)-\mathrm{Tr}(\rho^{1-\alpha}K_i^\dag \rho^{\alpha}K_i)-\mathrm{Tr}(\rho^{1-\beta}K_i^\dag \rho^{\beta}K_i)].
$$
If $[\rho^{\alpha+\beta},K_i]=0$, the above equation becomes
$$
I^{\alpha,\beta}(\rho,K_i)=\frac{1}{4}[2\mathrm{Tr}(\rho K_i^\dag K_i)-\mathrm{Tr}(\rho^{1-\alpha}K_i^\dag \rho^{\alpha}K_i)-\mathrm{Tr}(\rho^{1-\beta}K_i^\dag \rho^{\beta}K_i)].
$$
For $p\in (0,1)$, $f(x)=x^p$ is operator monotone as well as
operator concave \cite{RB}. So in either case, one can prove the
conclusion (iv) by following the same line of the proof of Theorem 4
in \cite{LWJ} under the assumptions of item (iv).

At last, since (iii) holds, imitating the proof of Theorem 4 in \cite{LWJ}, one proves that (v) is also true. $\Box$

{\bf Remark 2} By Lemma 3 in \cite{LWJ}, it is observed that
$\mathcal{E}^\dag(K_i)=K_i$, $\mathcal{E}^\dag(K_i^\dag
K_i)=K_i^\dag K_i$ and $\mathcal{E}^\dag(K_i K_i^\dag)=K_i K_i^\dag$
if and only if $[E_i,K_j]=0$ and $[E_i,K_j^\dag]=0$ for all $i$ and
$j$. The later one could be seen as the condition that the channel
$\mathcal{E}$ does not disturb channel $\Phi$, which was proposed as
the assumption in proving the monotonicity of $I(\rho,\Phi)$ in
\cite{LUO11}. Taking this fact into consideration, it follows from
Theorem 1 that $I^{\alpha,\beta}(\rho,\Phi)$ $(\alpha,\beta\geq
0,~\alpha+\beta\leq 1)$ could be viewed as a family of coherence
measures with respect to a channel $\Phi$ under the restrictive
conditions.

{\bf Remark 3} Direct computation shows that property (iv) (ancillary independence) in Theorem 1 does not hold for $V^{\alpha,\beta}(\rho,\Phi)$ $(\alpha,\beta\geq 0,~\alpha+\beta\leq 1)$ unless $\alpha=\beta$. But the proof of strong monotonicity in Theorem 1 relies heavily on the property of the ancillary independence. Hence the method used in \cite{LWJ} fails, and we do not know whether $V^{\alpha,\beta}(\rho,\Phi)$ can be regarded as a family of coherence measures with respect to a channel in general. Note that $V^{\alpha,\alpha}(\rho,\Phi)=I^{\alpha,\alpha}(\rho,\Phi)$.
$V^{\alpha,\alpha}(\rho,\Phi)$ $(0\leq \alpha \leq 1)$ is just a special class of $I^{\alpha,\beta}(\rho,\Phi)$ $(\alpha,\beta\geq 0,~\alpha+\beta\leq 1)$.

Now we calculate $I^{\alpha,\beta}(\rho,\Phi)$ and $V^{\alpha,\beta}(\rho,\Phi)$ for qubit states with respect to different kinds of quantum channels.
Consider the Pauli channel defined by
\begin{equation}\label{eq23}
\Phi(\rho)=\sum_{j=0}^3 p_j\sigma_j\rho\sigma_j,~\,\,p_j\geq 0,\,\,~\sum_{j=0}^3p_j=1,
\end{equation}
where $\sigma_0=I$, and $\sigma_i$, $i=1,2,3$, are Pauli matrices.
For a qubit state
$\rho=\frac{1}{2}(\mathbf{1}+\mathbf{r}\cdot\mathbf{\sigma})$, where
$\mathbf{r}=\{r_1,r_2,r_3\}$ and
$\mathbf{\sigma}=\{\sigma_1,\sigma_2,\sigma_3\}$, its eigenvalues
are $\lambda_{1,2}=(1\mp |\mathbf{r}|)/2$, and
$$\rho^{\alpha}=\left(\begin{array}{cc}
         \frac{\lambda_1^{\alpha}+\lambda_2^{\alpha}}{2}+\frac{r_3(\lambda_2^{\alpha}-\lambda_1^{\alpha})}{2|\mathbf{r}|}&\frac{(-r_1+ir_2)(\lambda_1^{\alpha}-\lambda_2^{\alpha})}{2|\mathbf{r}|}\\
         \frac{(-r_1-ir_2)(\lambda_1^{\alpha}-\lambda_2^{\alpha})}{2|\mathbf{r}|}&\frac{\lambda_1^{\alpha}+\lambda_2^{\alpha}}{2}-\frac{r_3(\lambda_2^{\alpha}-\lambda_1^{\alpha})}{2|\mathbf{r}|}\\
         \end{array}
         \right).
$$
Then we have
\begin{equation}\label{eq24}
I^{\alpha,\beta}(\rho,\Phi)
=\frac{1}{4}\sum_{j=1}^3 p_j\frac{|\mathbf{r}|^2-r_j^2}{|\mathbf{r}|^2}
(\lambda_1^{\alpha}-\lambda_2^{\alpha})(\lambda_1^{\beta}-\lambda_2^{\beta})(\lambda_1^{1-\alpha-\beta}+\lambda_2^{1-\alpha-\beta}).
\end{equation}
In particular, when $p_1=p_2=p_3=p$, the Pauli channel defined in
Eq. (\ref{eq23}) becomes the depolarizing channel, and we have
\begin{equation}\label{eq25}
I^{\alpha,\beta}(\rho,\Phi)
=\frac{1}{2}p
(\lambda_1^{\alpha}-\lambda_2^{\alpha})(\lambda_1^{\beta}-\lambda_2^{\beta})(\lambda_1^{1-\alpha-\beta}+\lambda_2^{1-\alpha-\beta}),
\end{equation}
which is an increasing function of $p$. Taking $p_1=p$ and
$p_2=p_3=0$ in Eq. (\ref{eq23}), we get the bit-flipping channel,
and have
\begin{equation}\label{eq26}
I^{\alpha,\beta}(\rho,\Phi)
=\frac{1}{4}\cdot\frac{p(r_2^2+r_3^2)}{|\mathbf{r}|^2}
(\lambda_1^{\alpha}-\lambda_2^{\alpha})(\lambda_1^{\beta}-\lambda_2^{\beta})(\lambda_1^{1-\alpha-\beta}+\lambda_2^{1-\alpha-\beta}).
\end{equation}
Setting $p_1=p_2=0$ and $p_3=p$ in Eq. (\ref{eq23}), we obtain the
phase-flipping channel, and get
\begin{equation}\label{eq27}
I^{\alpha,\beta}(\rho,\Phi)
=\frac{1}{4}\cdot\frac{p(r_1^2+r_2^2)}{|\mathbf{r}|^2}
(\lambda_1^{\alpha}-\lambda_2^{\alpha})(\lambda_1^{\beta}-\lambda_2^{\beta})(\lambda_1^{1-\alpha-\beta}+\lambda_2^{1-\alpha-\beta}).
\end{equation}

For the (unital) amplitude damping channel $\Phi(\rho)=\sum_{j=1}^2 K_j\rho K_j^\dag$ with
\begin{equation}\label{ut}
K_1=|0\rangle\langle0|+\sqrt{1-q}|1\rangle\langle 1|,~~K_2=\sqrt{q}|1\rangle\langle 1|,~~0\leq q\leq 1,
\end{equation}
we have
\begin{equation}\label{eq29}
I^{\alpha,\beta}(\rho,\Phi)
=\frac{1}{4}\cdot\frac{(1-\sqrt{1-q})(r_1^2+r_2^2)}{2|\mathbf{r}|^2}
(\lambda_1^{\alpha}-\lambda_2^{\alpha})(\lambda_1^{\beta}-\lambda_2^{\beta})(\lambda_1^{1-\alpha-\beta}+\lambda_2^{1-\alpha-\beta}).
\end{equation}
And for the (nonunital) amplitude damping channel
$\Phi(\rho)=\sum_{j=1}^2 K_j\rho K_j^\dag$ with
\begin{equation}\label{nut}
K_1=|0\rangle\langle0|+\sqrt{1-q}|1\rangle\langle 1|,~~K_2=\sqrt{q}|0\rangle\langle 1|,~~0\leq q\leq 1,
\end{equation}
we have
\begin{eqnarray}\label{eq31}
I^{\alpha,\beta}(\rho,\Phi)
&=&\frac{1}{4}\left[\frac{(1-\sqrt{1-q})(r_1^2+r_2^2)+qr_3^2}{2|\mathbf{r}|^2}(\lambda_1^{1-\alpha-\beta}+\lambda_2^{1-\alpha-\beta})\right.
\nonumber\\
&&\left.+\frac{qr_3}{2|\mathbf{r}|}(\lambda_1^{1-\alpha-\beta}-\lambda_2^{1-\alpha-\beta})\right]
(\lambda_1^{\alpha}-\lambda_2^{\alpha})(\lambda_1^{\beta}-\lambda_2^{\beta}).
\end{eqnarray}

In a similar way, we can also compute $V^{\alpha,\beta}(\rho,\Phi)$
for a qubit state
$\rho=\frac{1}{2}(I+\mathbf{r}\cdot\mathbf{\sigma})$ with respect to
different channels. For the Pauli channel $\Phi$ defined in Eq.
(\ref{eq23}), we have
\begin{equation}\label{eq32}
V^{\alpha,\beta}(\rho,\Phi)
=\frac{1}{4}\sum_{j=1}^3 p_j\frac{|\mathbf{r}|^2-r_j^2}{4|\mathbf{r}|^2}
(\lambda_1^{\alpha}-\lambda_2^{\alpha}+\lambda_1^{\beta}-\lambda_2^{\beta})^2(\lambda_1^{1-\alpha-\beta}+\lambda_2^{1-\alpha-\beta}).
\end{equation}
For the depolarizing channel $\Phi$ (a special case when $p_1=p_2=p_3=p$), we have
\begin{equation}\label{eq33}
V^{\alpha,\beta}(\rho,\Phi)
=\frac{1}{8}p
(\lambda_1^{\alpha}-\lambda_2^{\alpha}+\lambda_1^{\beta}-\lambda_2^{\beta})^2(\lambda_1^{1-\alpha-\beta}+\lambda_2^{1-\alpha-\beta}),
\end{equation}
which is also an increasing function of $p$. For the bit-flipping
channel $\Phi$ (a special case when $p_1=p$ and $p_2=p_3=0$), we
have
\begin{equation}\label{eq34}
V^{\alpha,\beta}(\rho,\Phi)
=\frac{1}{4}\cdot\frac{p(r_2^2+r_3^2)}{4|\mathbf{r}|^2}
(\lambda_1^{\alpha}-\lambda_2^{\alpha}+\lambda_1^{\beta}-\lambda_2^{\beta})^2(\lambda_1^{1-\alpha-\beta}+\lambda_2^{1-\alpha-\beta}).
\end{equation}
For the phase-flipping channel $\Phi$ (a special case when
$p_1=p_2=0$ and $p_3=p$), we have
\begin{equation}\label{eq35}
V^{\alpha,\beta}(\rho,\Phi)
=\frac{1}{4}\cdot\frac{p(r_1^2+r_2^2)}{4|\mathbf{r}|^2}
(\lambda_1^{\alpha}-\lambda_2^{\alpha}+\lambda_1^{\beta}-\lambda_2^{\beta})^2(\lambda_1^{1-\alpha-\beta}+\lambda_2^{1-\alpha-\beta}).
\end{equation}

For the (unital) amplitude damping channel (\ref{ut}), we have
\begin{equation}\label{eq36}
V^{\alpha,\beta}(\rho,\Phi)
=\frac{1}{4}\cdot\frac{(1-\sqrt{1-q})(r_1^2+r_2^2)}{8|\mathbf{r}|^2}
(\lambda_1^{\alpha}-\lambda_2^{\alpha}+\lambda_1^{\beta}-\lambda_2^{\beta})^2(\lambda_1^{1-\alpha-\beta}+\lambda_2^{1-\alpha-\beta}).
\end{equation}
And for the (nonunital) amplitude damping channel (\ref{nut}), we
have
\begin{eqnarray}\label{eq37}
V^{\alpha,\beta}(\rho,\Phi)
&=&\frac{1}{4}\left[\frac{(1-\sqrt{1-q})(r_1^2+r_2^2)+qr_3^2}{8|\mathbf{r}|^2}(\lambda_1^{1-\alpha-\beta}+\lambda_2^{1-\alpha-\beta})\right.
\nonumber\\
&&\left.+\frac{qr_3}{8|\mathbf{r}|}(\lambda_1^{1-\alpha-\beta}-\lambda_2^{1-\alpha-\beta})\right]
(\lambda_1^{\alpha}-\lambda_2^{\alpha}+\lambda_1^{\beta}-\lambda_2^{\beta})^2.
\end{eqnarray}
We can see that when $\alpha=\beta=\frac{1}{2}$, the above
analytical expressions for a qubit state with respect to certain
channels reduce to the corresponding ones in Ref. \cite{LUO11}.

\vskip0.1in

\noindent {\bf 5. Complementarity relations and information
conservation}

\vskip0.1in

In Section 3, we have defined $I^{\alpha,\beta}(\rho,\Phi)$ and
$J^{\alpha,\beta}(\rho,\Phi)$ using commutator and anti-commutator,
which is in a dual fashion in some sense. Another kind of
generalization of the quantities $V^{\alpha,\beta}(\rho,\Phi)$ and
$W^{\alpha,\beta}(\rho,\Phi)$ have also been introduced. In this
section, we discuss the relationships between these two sets of
quantities.

Set
$$
C^{\alpha,\beta}(\rho,K)=\frac{1}{2}[\mathrm{Tr}(\rho^{1-\alpha}K^\dag \rho^{\alpha}K)+\mathrm{Tr}(\rho^{1-\beta}K^\dag \rho^{\beta}K)],~~
\alpha,\beta\geq 0,~\alpha+\beta\leq 1.
$$
It follows from Eqs. (\ref{eq15}) and (\ref{eq16}) that
$$
I^{\alpha,\beta}(\rho,K)=\frac{1}{4}[\mathrm{Tr}(\rho^{1-\alpha-\beta}K^\dag \rho^{\alpha+\beta}K+K\rho K^\dag)-2C^{\alpha,\beta}(\rho,K)]
$$
and
$$
J^{\alpha,\beta}(\rho,K)=\frac{1}{4}[\mathrm{Tr}(\rho^{1-\alpha-\beta}K^\dag \rho^{\alpha+\beta}K+K\rho K^\dag)+2C^{\alpha,\beta}(\rho,K)].
$$
Noting that
$$C^{\alpha,\beta}(\rho,K)=\frac{1}{2}(\|\rho^{\frac{\alpha}{2}}K\rho^{\frac{1-\alpha}{2}}\|_2^2
+\|\rho^{\frac{\beta}{2}}K\rho^{\frac{1-\beta}{2}}\|_2^2)\geq 0,$$
we have
$$I^{\alpha,\beta}(\rho,K)\leq J^{\alpha,\beta}(\rho,K),$$
and thus
$$I^{\alpha,\beta}(\rho,\Phi)\leq J^{\alpha,\beta}(\rho,\Phi).$$

For any channel $\Phi$ with Kraus representation in the form of Eq.
(\ref{eq10}), the quantities $I^{\alpha,\beta}(\rho,\Phi)$ and
$J^{\alpha,\beta}(\rho,\Phi)$ can be rewritten as
$$
I^{\alpha,\beta}(\rho,\Phi)=\frac{1}{4}\mathrm{Tr}[\rho^{1-\alpha-\beta}\Phi^\dag(\rho^{\alpha+\beta})+\Phi(\rho)
-\rho^{1-\alpha}\Phi^\dag(\rho^{\alpha})-\rho^{1-\beta}\Phi^\dag(\rho^{\beta})]
$$
and
$$
J^{\alpha,\beta}(\rho,\Phi)=\frac{1}{4}\mathrm{Tr}[\rho^{1-\alpha-\beta}\Phi^\dag(\rho^{\alpha+\beta})+\Phi(\rho)
+\rho^{1-\alpha}\Phi^\dag(\rho^{\alpha})+\rho^{1-\beta}\Phi^\dag(\rho^{\beta})],
$$
which implies that for $\alpha,\beta\geq 0,~\alpha+\beta\leq 1$,
\begin{equation}\label{eq38}
I^{\alpha,\beta}(\rho,\Phi)+J^{\alpha,\beta}(\rho,\Phi)=\frac{1}{2}\mathrm{Tr}(\rho^{1-\alpha-\beta}\Phi^\dag(\rho^{\alpha+\beta})+\Phi(\rho)).
\end{equation}

When $\alpha+\beta=1$, Eq. (\ref{eq38}) reduces to the following
one:
\begin{equation}\label{eq39}
I^{\alpha}(\rho,\Phi)+J^{\alpha}(\rho,\Phi)=\frac{1}{2}\mathrm{Tr}[\Phi^\dag(\rho)+\Phi(\rho)],~~0\leq \alpha \leq 1.
\end{equation}
In particular, when $\Phi$ is a unital quantum operation, i.e.,
$\sum_{i}K_i^\dag K_i=I$ and $\sum_{i}K_i K_i^\dag=I$, we get
\begin{equation}\label{eq40}
I^{\alpha}(\rho,\Phi)+J^{\alpha}(\rho,\Phi)=1,~~0\leq \alpha \leq 1.
\end{equation}
which could be viewed as a class of conservation relations.

%Note that when $\alpha=\frac{1}{2}$, Eqs. (\ref{eq28}) and (\ref{eq29}) reduce to Eqs. (32) and (33) in Ref. \cite{LUO11}.

In a similar manner, we can investigate the relationship between $V^{\alpha,\beta}(\rho,\Phi)$ and $W^{\alpha,\beta}(\rho,\Phi)$. Define the following quantity
$$
D^{\alpha,\beta}(\rho,K)=\mathrm{Tr}\left(\frac{\rho^{1-\alpha}+\rho^{1-\beta}}{2}K^\dag
\frac{\rho^{\alpha}+\rho^{\beta}}{2}K\right),~~ \alpha,\beta\geq
0,~\alpha+\beta\leq 1.
$$
It follows from Eqs. (\ref{eq19}) and (\ref{eq20}) that
\begin{eqnarray*}
V^{\alpha,\beta}(\rho,K)
&=&\frac{1}{4}\left[\mathrm{Tr}\left(\rho^{1-\alpha-\beta}K^\dag
\left(\frac{\rho^{\alpha}+\rho^{\beta}}{2}\right)^2
K+K\left(\frac{\rho^{1-\alpha}+\rho^{1-\beta}}{2}\right)\left(\frac{\rho^{\alpha}+\rho^{\beta}}{2}\right)K^\dag\right)\right.
\nonumber\\
&&\left.-2D^{\alpha,\beta}(\rho,K)\right]
\end{eqnarray*}
and
\begin{eqnarray*}
W^{\alpha,\beta}(\rho,K)
&=&\frac{1}{4}\left[\mathrm{Tr}\left(\rho^{1-\alpha-\beta}K^\dag
\left(\frac{\rho^{\alpha}+\rho^{\beta}}{2}\right)^2
K+K\left(\frac{\rho^{1-\alpha}+\rho^{1-\beta}}{2}\right)\left(\frac{\rho^{\alpha}+\rho^{\beta}}{2}\right)K^\dag\right)\right.
\nonumber\\
&&\left.+2D^{\alpha,\beta}(\rho,K)\right].
\end{eqnarray*}
Noting that
$$D^{\alpha,\beta}(\rho,K)=\frac{1}{4}(\|\rho^{\frac{\alpha}{2}}K\rho^{\frac{1-\alpha}{2}}\|_2^2
+\|\rho^{\frac{\beta}{2}}K\rho^{\frac{1-\beta}{2}}\|_2^2
+\rho^{\frac{\alpha}{2}}K\rho^{\frac{1-\beta}{2}}\|_2^2
+\rho^{\frac{\beta}{2}}K\rho^{\frac{1-\alpha}{2}}\|_2^2)
\geq 0,$$
we have
$$V^{\alpha,\beta}(\rho,K)\leq W^{\alpha,\beta}(\rho,K),$$
and thus
$$V^{\alpha,\beta}(\rho,\Phi)\leq W^{\alpha,\beta}(\rho,\Phi).$$

For any channel $\Phi$ with Kraus representation in the form of Eq.
(\ref{eq10}), the quantities $V^{\alpha,\beta}(\rho,\Phi)$ and
$W^{\alpha,\beta}(\rho,\Phi)$ can be rewritten as
\begin{eqnarray*}
V^{\alpha,\beta}(\rho,\Phi)
&=&\frac{1}{4}\mathrm{Tr}\left[\rho^{1-\alpha-\beta}\Phi^\dag\left(\left(\frac{\rho^{\alpha}+\rho^{\beta}}{2}\right)^2\right)
+\Phi\left(\left(\frac{\rho^{1-\alpha}+\rho^{1-\beta}}{2}\right)\left(\frac{\rho^{\alpha}+\rho^{\beta}}{2}\right)\right)\right.
\nonumber\\
&&\left.-2\left(\frac{\rho^{1-\alpha}+\rho^{1-\beta}}{2}\right)\Phi^\dag(\frac{\rho^{\alpha}+\rho^{\beta}}{2})\right]
\end{eqnarray*}
and
\begin{eqnarray*}
W^{\alpha,\beta}(\rho,\Phi)
&=&\frac{1}{4}\mathrm{Tr}\left[\rho^{1-\alpha-\beta}\Phi^\dag\left(\left(\frac{\rho^{\alpha}+\rho^{\beta}}{2}\right)^2\right)
+\Phi\left(\left(\frac{\rho^{1-\alpha}+\rho^{1-\beta}}{2}\right)\left(\frac{\rho^{\alpha}+\rho^{\beta}}{2}\right)\right)\right.
\nonumber\\
&&\left.+2\left(\frac{\rho^{1-\alpha}+\rho^{1-\beta}}{2}\right)
\Phi^\dag\left(\frac{\rho^{\alpha}+\rho^{\beta}}{2}\right)\right],
\end{eqnarray*}
which implies that for $\alpha,\beta\geq 0$, $\alpha+\beta\leq 1$,
\begin{eqnarray}\label{eq41}
&&V^{\alpha,\beta}(\rho,\Phi)+W^{\alpha,\beta}(\rho,\Phi)\nonumber\\
&&=\frac{1}{4}\mathrm{Tr}\left[\rho^{1-\alpha-\beta}\Phi^\dag\left(\left(\frac{\rho^{\alpha}+\rho^{\beta}}{2}\right)^2\right)
+\Phi\left(\left(\frac{\rho^{1-\alpha}+\rho^{1-\beta}}{2}\right)\left(\frac{\rho^{\alpha}+\rho^{\beta}}{2}\right)\right)\right].
\end{eqnarray}

Furthermore, when $\alpha=\beta=\frac{1}{2}$, Eq. (\ref{eq41})
reduces to
\begin{equation}\label{eq42}
V^{\frac{1}{2},\frac{1}{2}}(\rho,\Phi)+W^{\frac{1}{2},\frac{1}{2}}(\rho,
\Phi)=\frac{1}{2}\mathrm{Tr}[\Phi^\dag(\rho)+\Phi(\rho)].
\end{equation}
In particular, when $\Phi$ is a unital quantum operation, i.e.,
$\sum_{i}K_i^\dag K_i=I$ and $\sum_{i}K_i K_i^\dag=I$, we obtain
\begin{equation}\label{eq43}
V^{\frac{1}{2},\frac{1}{2}}(\rho,\Phi)+W^{\frac{1}{2},\frac{1}{2}}(\rho,
\Phi)=1,
\end{equation}
which could also be regarded as a conservation relation. Note that
from the definitions,
$V^{\frac{1}{2},\frac{1}{2}}(\rho,K)=I(\rho,K)$ and
$W^{\frac{1}{2},\frac{1}{2}}(\rho,K)=J(\rho,K)$, and thus
$V^{\frac{1}{2},\frac{1}{2}}(\rho,\Phi)=I(\rho,\Phi)$ and
$W^{\frac{1}{2},\frac{1}{2}}(\rho,\Phi)=J(\rho,\Phi)$, Eqs.
(\ref{eq42}) and (\ref{eq43}) hold naturally since they are in
accord with Eqs. (32) and (33) in Ref. \cite{LUO11}, respectively.

Inspired by Ref. \cite{LUO11}, we now give two interpretations of
the conservation relations Eq. (\ref{eq40}) derived above. First of
all, for any group representation $\{U(g):g\in G\}$ on a finite
group $G$, we can define a quantum channel
\begin{equation}\label{eq44}
\Phi(\rho)=\frac{1}{|G|}\sum_{g\in G}U(g)\rho U(g)^\dag,
\end{equation}
where $|G|$ denotes the number of elements in $G$. For a compact Lie
group $G$, it can be also defined that
\begin{equation}\label{eq45}
\Phi(\rho)=\int \mathrm{d}\mu(g) U(g)\rho U(g)^\dag,
\end{equation}
where $U(g)$ is the unitary representation of $g\in G$ and
$\mathrm{d}\mu$ is the Haar measure with respect to $G$. In this
regard, it is natural to interpret $J^{\alpha}(\rho,\Phi)$ and
$I^{\alpha}(\rho,\Phi)$ as symmetry and asymmetry of $\rho$ with
respect to the group $G$, respectively. Therefore, Eq. (\ref{eq40})
could be viewed as a family of symmetry-asymmetry complementarity
relations.

Secondly, we illustrate Eq. (\ref{eq40}) with wave-particle duality
in the Mach-Zehnder interferometry with a path detector
\cite{SPEAE}. The beam-splitter, mirror and phase shift are realized
by the following unitary matrices
$$U_B=\frac{1}{\sqrt{2}}\left(\begin{array}{cc}
         1&1\\
         1&-1\\
         \end{array}
         \right),~~
U_M=\left(\begin{array}{cc}
         0&1\\
         1&0\\
         \end{array}
         \right),~~
U_{\theta}=\left(\begin{array}{cc}
         e^{i\theta}&0\\
         0&1\\
         \end{array}
         \right),
$$
respectively, while the path detector is realized by the unitary
matrix $V$. The input state is a bipartite state
$\rho^{ab}=\rho\otimes \tau$, where
$\rho=\frac{1}{2}(\mathbf{1}+\mathbf{r}\cdot\mathbf{\sigma})$ is the
initial (external) system state and $\tau$ is the initial (internal)
state. Here $\mathbf{r}=\{r_1,r_2,r_3\}$ is the Bloch vector with
$|\mathbf{r}|^2\leq 1$ and $\mathbf{\sigma}=\{\sigma_1,\sigma_2,\sigma_3\}$
with $\sigma_i~(i=1,2,3)$ the Pauli matrices. The interferometry
channel is
\begin{equation}\label{eq46}
\Phi(\rho)=\mathrm{Tr}_{b}(U(\rho\otimes \tau)U^\dag),
\end{equation}
where $U=U_{B}^{ab}U_{M}^{ab}V^{ab}U_{B}^{ab}$,
$U_{B}^{ab}=U_{B}\otimes \mathbf{1}^{b}$, $U_{M}^{ab}=U_{M}\otimes
\mathbf{1}^{b}$ and $V^{ab}=e^{i\theta}|0\rangle\langle 0|\otimes
\mathbf{1}^{b}+|1\rangle\langle 1|\otimes V.$ Direct calculation
shows that
\begin{eqnarray}\label{eq47}
I^{\alpha}(\rho,\Phi)
&=&\frac{1}{4}\left[2-(\lambda_1^{1-\alpha}+\lambda_2^{1-\alpha})(\lambda_1^{\alpha}+\lambda_2^{\alpha})
+\frac{(\lambda_1^{1-\alpha}-\lambda_2^{1-\alpha})(\lambda_1^{\alpha}-\lambda_2^{\alpha})}{|r|^2}r_1^2\right.
\nonumber\\
&&\left.-\frac{(\lambda_1^{1-\alpha}-\lambda_2^{1-\alpha})(\lambda_1^{\alpha}-\lambda_2^{\alpha})}{|r|^2}(1-r_1^2)|\mathrm{Tr}(V\tau)|\mathrm{cos}(\theta-\nu-\gamma)
\right],
\end{eqnarray}
and
\begin{eqnarray}\label{eq48}
J^{\alpha}(\rho,\Phi)
&=&\frac{1}{4}\left[2+(\lambda_1^{1-\alpha}+\lambda_2^{1-\alpha})(\lambda_1^{\alpha}+\lambda_2^{\alpha})
-\frac{(\lambda_1^{1-\alpha}-\lambda_2^{1-\alpha})(\lambda_1^{\alpha}-\lambda_2^{\alpha})}{|r|^2}r_1^2\right.
\nonumber\\
&&\left.+\frac{(\lambda_1^{1-\alpha}-\lambda_2^{1-\alpha})(\lambda_1^{\alpha}-\lambda_2^{\alpha})}{|r|^2}(1-r_1^2)|\mathrm{Tr}(V\tau)|\mathrm{cos}(\theta-\nu-\gamma)
\right],
\end{eqnarray}
where $\nu=\mathrm{arg}(\mathrm{Tr}(V\tau))$,
$\gamma=\mathrm{arctan}\frac{2r_2r_3}{r_2^2-r_3^2}$ and
$\lambda_{1,2}=(1\mp |\mathbf{r}|)/2$ are the eigenvalues of the
qubit state
$\rho=\frac{1}{2}(\mathbf{1}+\mathbf{r}\cdot\mathbf{\sigma})$.

Since $I^{\alpha}(\rho,\Phi)$ is a manifestation of the asymmetry,
by minimizing over the phase shift, the quantity
\begin{eqnarray}\label{eq49}
\tilde{P}^{\alpha}(\rho,\Phi)&=&\min_{\theta}I^{\alpha}(\rho,\Phi)
\nonumber\\
&=&\frac{1}{4}\left[2-(\lambda_1^{1-\alpha}+\lambda_2^{1-\alpha})(\lambda_1^{\alpha}+\lambda_2^{\alpha})
+\frac{(\lambda_1^{1-\alpha}-\lambda_2^{1-\alpha})(\lambda_1^{\alpha}-\lambda_2^{\alpha})}{|r|^2}r_1^2\right.
\nonumber\\
&&\left.-\frac{(\lambda_1^{1-\alpha}-\lambda_2^{1-\alpha})(\lambda_1^{\alpha}-\lambda_2^{\alpha})}{|r|^2}(1-r_1^2)|\mathrm{Tr}(V\tau)|
\right]
\end{eqnarray}
gives a quantification for the which-path information. In contrast,
since $J^{\alpha}(\rho,\Phi)$ represents the symmetry, by maximizing
over the phase shift, the quantity
\begin{eqnarray}\label{eq50}
\tilde{W}^{\alpha}(\rho,\Phi)&=&\max_{\theta}J^{\alpha}(\rho,\Phi)
\nonumber\\
&=&\frac{1}{4}\left[2+(\lambda_1^{1-\alpha}+\lambda_2^{1-\alpha})(\lambda_1^{\alpha}+\lambda_2^{\alpha})
-\frac{(\lambda_1^{1-\alpha}-\lambda_2^{1-\alpha})(\lambda_1^{\alpha}-\lambda_2^{\alpha})}{|r|^2}r_1^2\right.
\nonumber\\
&&\left.+\frac{(\lambda_1^{1-\alpha}-\lambda_2^{1-\alpha})(\lambda_1^{\alpha}-\lambda_2^{\alpha})}{|r|^2}(1-r_1^2)|\mathrm{Tr}(V\tau)|
\right]
\end{eqnarray}
could be exploited as a measure of the interference visibility.
Utilizing these quantities with parameter $\alpha$, we thus obtain a
family of complementarity relations signifying the wave-particle
duality:
\begin{equation}
\tilde{P}^{\alpha}(\rho,\Phi)+\tilde{W}^{\alpha}(\rho,\Phi)=1,\,\,\,\,\,0\leq
\alpha\leq 1,
\end{equation}
which is more general than Eq. (\ref{eq43}) in Ref. \cite{LUO11}.

\vskip0.1in

\noindent {\bf 6. Conclusions and discussions}

\vskip0.1in

We have showed that the state-channel interactions based on modified
generalized Wigner-Yanase-Dyson skew information could be exploited
as a family of coherence measures with respect to a quantum channel
under certain assumptions. We have also provided explicit analytical
expressions of this family of coherence measures for qubit states
with respect to various kinds of quantum channels. Moreover,
complementarity relations based on MGWYD skew information and MWGWYD
skew information have been derived. Inspired by Ref.\cite{LUO11}, we
have also presented two interpretations of the conservation relations:
the interplay between symmetry and asymmetry with
respect to a group, and the wave-particle duality.

Coherence and complementarity are both fundamental issues in quantum
mechanics and quantum information theory. As is shown in
Ref.\cite{LUO11}, the quantities $I(\rho,\Phi)$ and $J(\rho,\Phi)$
not only present a basic and an alternative framework for addressing
complementarity, but also put forward the study of coherence in a broad
context involving quantum channels. The quantity $I(\rho,\Phi)$ has several
interpretations including asymmetry, coherence, incompatibility,
quantumness and quantum uncertainty with state-channel interactions,
while $J(\rho,\Phi)$ possesses the corresponding interpretations in a dual
fashion. We have revisited the work of Refs.\cite{LUO11} and \cite{LWJ}
based on the interplay of $I^{\alpha,\beta}({\rho},\Phi)$ and
$J^{\alpha,\beta}({\rho},\Phi)$ as well as
$V^{\alpha,\beta}({\rho},\Phi)$ and $W^{\alpha,\beta}({\rho},\Phi)$
induced by MGWYD skew information and MWGWYD skew information, and investigated
the coherence and complementarity based on measures with
parameters $\alpha$ and $\beta$. Therefore, our results are valid for a
large family of quantities, which are more general than the ones
induced by MWY skew information and MWYD skew information proposed
in Ref.\cite{LUO11} and Ref.\cite{LWJ}, respectively.

The quantities in Eqs. (\ref{eq8}) and (\ref{eq9}) are defined in
terms of the Hilbert-Schmidt norm. A natural question arises: could
$I({\rho},\Phi)=\sum_{i}\|[\rho^{\frac{1}{2}},K_i]\|_p^2$ be
exploited as a measure of coherence with respect to a quantum
channel? Or more generally, could
$I({\rho},\Phi)=\sum_{i}|||[\rho^{\frac{1}{2}},K_i]|||^2$ be
utilized as a measure of coherence with respect to a quantum channel
for other unitarily invariant norms other than Schatten $p$-norms?
These questions deserve further investigations.

\vskip0.1in

\noindent

%=============================================================================%
\subsubsection*{Acknowledgements}
This work was supported by National Natural Science Foundation of
China (Grant Nos. 11701259, 11971140, 11461045, 11675113, 11301124),
the China Scholarship Council (Grant No.201806825038), the Key
Project of Beijing Municipal Commission of Education under No.
KZ201810028042, Beijing Natural Science Foundation (Z190005),
Natural Science Foundation of Zhejiang Province of China
(LY17A010027). This work was completed while Zhaoqi Wu and Lin Zhang
were visiting Max-Planck-Institute for Mathematics in the Sciences
in Germany.

%===========================================================================%

\end{document}